\documentclass[a4paper]{article}
\def\be{\begin{equation}}
\def\eea{\end{eqnarray}}
\def\bea{\begin{eqnarray}}
\def\ee{\end{equation}}
\def\l{\lambda}
\def\r{\ref}
\def\o{\over}
\def\m{\mu}
\def\d{\mathrm{d}}
\def\a{\alpha}
\def\b{\beta}

\def\s{\sum}
\usepackage[psamsfonts]{amssymb}
\usepackage{amsmath}
\usepackage{cite}
\author{Masoud Alimohammadi\footnote{alimohmd@ut.ac.ir}
\\ {\small School of Physics, University of Tehran,}
\\ {\small North Karegar Ave., Tehran, Iran.}}
\title{ Solvable reaction-diffusion processes without exclusion}
\begin{document}
\maketitle
\begin{abstract}

\noindent For reaction-diffusion processes without exclusion, in
which the particles can exist in the same site of a
one-dimensional lattice, we study all the integrable models which
can be obtained by imposing a boundary condition on the master
equation of the asymmetric diffusion process. The annihilation
process is also added. The Bethe ansatz solution and the exact
$N$-particle conditional probabilities are obtained.
\end{abstract}
\section{Introduction}
There are a variety of phenomena that can be explained by
stochastic models, and their non-equilibrium behaviors can be
understood by rather simple rules \cite{1,2,3}. One of the
important examples of these models are the reaction-diffusion
processes on a one-dimensional lattice for which their dynamics
are fully specified by their master equation \cite {4,5}. A simple
example of reaction-diffusion process is asymmetric simple
exclusion process (ASEP) \cite{2,6,7}, which is known to be
relevant to various fields of science like the kinetics of
biopolymerization \cite{8,9}, traffic models \cite{10}, polymers
in random media, dynamical models of interface growth
\cite{11,12}, noisy Burgers equation \cite{13}, study of shocks
\cite {14,15}, sequence alignment \cite{116}, and molecular motors
\cite{117}. For recent reviews, see for instance \cite{16,17,18}.

The ASEP is a lattice model in which each particle hops to its
right (left) nearest-neighboring site with a probability $D_R\d t
(D_L\d t)$ in an infinitesimal interval $\d t$. In addition,
particles are subject to hard-core exclusion: each site is either
occupied only by one particle or empty. ASEP has been studied in
\cite{19} by introducing a master equation which describes the
time evolution of probabilities $P(x_1,\cdots ,x_N;t)$, when the
particles are not in neighboring sites, and a so-called boundary
condition, which specifies the situations in which the
probabilities go outside the physical region $x_1<x_2<\cdots
<x_N$. These happen when some of the particles are in adjacent
sites and the master equation can not be applied to them. It has
been shown that the model is integrable in the sense that the
$N$-particle $S$-matrix is factorized into a product of
two-particle $S$-matrices. The coordinate Bethe ansatz has been
used in this proof. Note that the $S$-matrix can completely
determine the dynamics of a Markovian process, i.e. the
$N$-particle probabilities of a model.

By choosing other suitable boundary conditions, without changing
the master equation, one may study the more complicated
reaction-diffusion processes, even with long-range interaction. In
ref.\cite{20}, the so-called drop-push model has been studied by
this method. In this model the particle hops to the next right
site even it is occupied. The particle hops to this site by
pushing all the neighboring particles to their next right sites,
with a rate depending on the number of right neighboring
particles. The generalization of this model, by considering both
the right and left hopping, has been done in ref.\cite{21}. This
method has been also applied to more-than-one-species situations,
which become more complicated. The complexity arises from the
above mentioned factorization of $N$-particle scattering matrix.
In these cases, the factorization demands the two-particle
$S$-matrices to satisfy the spectral Yang-Baxter equation. Various
solvable multi-species models have been studied in this way, of
which the most recent general cases have been discussed in
\cite{22} and \cite{23}.

All of the above studies have been restricted to interactions
which include the hard-core exclusion. This made some
simplification. In ref.\cite{24,25}, an asymmetric diffusion model
without exclusion has been shown to be integrable and to have the
same $R$-matrix as that of the ASEP. In \cite{26}, it has been
shown that the processes:
  \bea\label{1}
  mn &\rightarrow &m-l,n+l \
\ \ \ {\rm with
 \ rate}\ \ {D_R\over [l]},\cr
mn &\rightarrow &m+l,n-l \ \ \ \ {\rm with
 \ rate}\ \ {D_R\gamma^l\over [l]},
    \eea
is integrable, in the sense of the above mentioned two-particle
factorization. The numbers $"m", "n",\cdots$ indicate the particle
numbers on a site, $\gamma =D_L/D_R$, and
 \be\label{2}
 [l]={1-\gamma^l\over 1-\gamma}.
  \ee
Note that in \cite{26}, the reaction rates of eq.(\ref{1}) are
scaled by $D_R$. These processes are obtained by imposing the
boundary condition
 \be\label{3}
 P_N(\cdots , x_j,x_j-1,\cdots ;t)=D_RP_N(\cdots , x_j-1,x_j-1,\cdots ;t)
+D_LP_N(\cdots , x_j,x_j,\cdots ;t) \ \ \ (j=1,\cdots , N-1),
 \ee
on master equation of ASEP:
 \bea\label{4}
 {\partial \over \partial
 t}P(x_1,\cdots,x_j,\cdots,x_N;t)&=&\sum_{j=1}^N
 D_RP(x_1,\cdots,x_{j-1},x_j-1,x_{j+1}\cdots,x_N;t)\cr
 &&+\sum_{j=1}^ND_LP(x_1,\cdots,x_{j-1},x_j+1,x_{j+1}\cdots,x_N;t)\cr
  && -NP(x_1,\cdots,x_j,\cdots,x_N;t),
  \eea
in which we have used a time scale so that $D_R+D_L=1$. In above
equation, $P(x_1,\cdots,x_N;t)$ is the probability for finding at
time $t$ the particles at sites $x_1,\cdots,x_N$. We take these
functions to define probabilities only in the physical region
$x_1\leq x_2\leq \cdots \leq x_N$. In fact, in the domain
$\Omega_N=x_1\leq x_2\leq \cdots \leq x_N\subset Z^N$, the
function $P$ is the probability defined above, whereas in
$Z^N/\Omega_N$ it is defined by the master equation (\ref{4}), but
it is not a probability.
 The master equation (\ref{4})
is only valid for
 \be
 x_i<x_{i+1},
 \ee
 since for $x_i=x_{i+1}$, there will be terms with $x_i=x_{i+1}+1$
in the right-hand side of eq.(\ref{4}), which is out of the
physical region. One can, however, assume that (\ref{4}) is valid
for all physical region $x_i\leq x_{i+1}$, and impose certain
boundary condition for $x_i=x_{i+1}$. The boundary condition
(\ref{3}) leads to interactions (\ref{1}).

In this paper we want to study all possible boundary conditions
for single-species systems and derive all the integrable
one-dimensional reaction-diffusion processes without exclusion
which can be obtained by this method. The scheme of the paper is
the following. In section 2, we show that there are two types of
boundary conditions when $D_R\neq 0$ and $D_L\neq 0$. The first is
one considered in \cite{26}, i.e. eq.(\ref{3}), which we call it
the {\bf type 1} model, and the second one is:
 \be\label{6}
 D_RP(x,x-1)+D_LP(x+1,x)=P(x,x).
 \ee
Here we have suppressed all the other coordinates for simplicity.
We show that the interactions of this {\bf type 2} model are
 \bea\label{7}
  mn &\rightarrow &m-1,n+1 \
\ \ \ {\rm with
 \ rate}\ \ D_R,\cr
mn &\rightarrow &m+1,n-1 \ \ \ \ {\rm with
 \ rate}\ \ D_L.
    \eea
This is the reactions (\ref{1}), restricted to $l=1$. In other
words, to have an integrable model, it is not necessary to have
the simultaneous hoppings of any number of particles from a common
site to the neighboring site ( as indicated in (\ref{1})), but the
one particle hoppings can also lead to integrable models. Note
that the reactions (\ref{7}) is not a subset of processes
(\ref{1}).

In the totally asymmetric case with $D_L=0$, we show in section 3
that there is a new boundary condition which is a linear
combination of boundary conditions (\ref{3}) and (\ref{6}), i.e.
  \be\label{8}
  P(x,x-1)=\l P(x,x)+\m P(x-1,x-1).
  \ee
It is shown that this {\bf type 3} model contains the reactions
 \be\label{9}
  mn \rightarrow m-l,n+l,
  \ee
with rates
  \be\label{10}
  r_l={1\over {1+{\l \over \m}+\cdots +\left( {\l \over
  \m}\right)^{l-1}}}.
  \ee
This is an interesting one-parameter family of interactions.

In section 4, we generalize the boundary conditions (\ref{3}),
(\ref{6}) and (\ref{8}) to include annihilation to the processes
(\ref{1}), (\ref{7}) and (\ref{9}), respectively. The resulting
models are rather involved. To be specific, we consider $D_L=0$
case of (\ref{3}), and show that the boundary condition
 \be\label{11}
 P(x,x-1)=\m P(x-1,x-1) \ \ \ (\m <1)
 \ee
describes the reactions
  \bea\label{12}
  mn &\rightarrow &m-l,n+l \
\ \ \ {\rm with
 \ rate}\ \ \m^{l-1},\cr
n &\rightarrow & \left\{
\begin{array}{c}
n-1 \\n-2\\ \vdots
\end{array}
\right.
 \ \ \ \ {\rm with \ total
 \ rates}\ \ n-\sum_{l=0}^{n-1}\m^l.
    \eea
The second reactions are the annihilation processes. We call this
model as the {\bf type 4} model. It must be mentioned that we can
not extend our investigation to include the creation processes.
The main reason is that if we do so, the evolution equation of
$n$-particle sector will not become closed and will depend on the
more-than-$n$-particle configurations.

The Bethe ansatz solution for different models is discussed in
section 5 and the exact $N$-point conditional probabilities of
type 4 model is obtained in section 6. Some interesting physical
quantities are also obtained. Finally we discuss the multi-species
extension of these processes in the last section and show that
this generalization is not possible for the reactions without
exclusion.
\section{Reactions with $D_L\neq 0$ and $D_R\neq 0$}
Consider the master equation (\ref{4}) for two particle sector,
when the particles are at site $x_1=x_2=x$:
 \be\label{13}
 {\dot P}(x,x)=D_RP(x-1,x)+D_RP(x,x-1)+D_LP(x+1,x)+D_LP(x,x+1)
 -2P(x,x).
 \ee
The second and third probabilities in the right-hand side of the
above equation are out of the physical region, and must be defined
through some boundary conditions. There are only two possibilities
which are consistent in more-than-two-particle sectors. The first
one is
  \be\label{14}
  \m' P(x,x-1)=D_RP(x-1,x-1)+D_LP(x,x),
 \ee
and the second one is
  \be\label{15}
  D_RP(x,x-1)+D_LP(x+1,x)=\l'P(x,x).
 \ee
In the first choice, we take any unphysical terms of eq.(\ref{13})
as a linear combination of physical functions, and in the second
choice, we take the whole unphysical terms as a linear combination
of the physical probabilities. The right-hand sides of
eqs.(\ref{14}) and (\ref{15}) are the only allowed combinations
which one can write. This was discussed in \cite{23} for ASEP
cases. In fact, one can use any other parameters $\alpha$ and
$\beta$ instead of $D_R$ and $D_L$ in eq.(\ref{14}), with
condition $\alpha +\beta=1$. Also one can add two other terms $\m
P(x-1,x-1)$ and $\nu P(x+1,x+1)$ to the right-hand side of
eq.(\ref{15}), but it can be shown that for obtaining a consistent
description in more-than-two-particle sectors, one must take $\mu
=\nu =0$ ( see section 3 of \cite{23} for more details in ASEP
case).

To obtain the range of parameters $\mu'$ and $\l'$, one can use
eq.(\ref{4}) to show that
  \bea\label{16}
  {\partial \o {\partial t}}\sum_{x_2}\s_{x_1\leq x_2}P(x_1,x_2;t)&=&
  -\s_xP(x,x;t)+D_R\s_xP(x,x-1;t)+D_L\s_xP(x+1,x;t) \cr &=&
  -\s_xP(x,x;t)+\s_xP(x,x-1;t).
  \eea
For boundary condition (\ref{14}), eq.(\ref{16}) results
  \be\label{17}
  {\partial \o {\partial t}}\s_{x_2}\s_{x_1\leq x_2}P(x_1,x_2;t)=
  \left( {1\o \m'}-1\right)\s_xP(x,x;t),
  \ee
and for boundary condition (\ref{15}), it results
   \be\label{18}
  {\partial \o {\partial t}}\s_{x_2}\s_{x_1\leq x_2}P(x_1,x_2;t)=
  ( \l'-1)\s_xP(x,x;t).
  \ee
In the first step, let us exclude the annihilation processes and
therefore it is assumed that the number of particles is constant
in time. In this way, the eqs.(\ref{17}) and (\ref{18}) lead to
  \bea\label{19}
  \m'=1&&  ({\rm for \ boundary \ condition\ (\ref{14})}),\cr
 \l'=1&&  ({\rm for \ boundary \ condition\ (\ref{15})}).
  \eea
Eq.(\ref{14}) exactly becomes the one studied in \cite{26}, i.e.
eq.(\ref{3}), and it induces the reactions (\ref{1}), but the
second one, eq.(\ref{15}), is new. To obtain the reaction
introduced by this boundary condition, with $\l'=1$, we first
consider ${\dot P(x,x)}$ in eq.(\ref{13}). Using eqs.(\ref{15})
and (\ref{19}), one finds
 \be\label{20}
 {\dot P}(x,x)=D_RP(x-1,x)+D_LP(x,x+1)
 -P(x,x),
 \ee
which is the evolution equation of the following two-particle
reactions
 \bea\label{21}
  10 &\rightarrow &01 \
\ \ \ {\rm with
 \ rate}\ \ D_R,\cr
01 &\rightarrow &10 \ \ \ \ {\rm with
 \ rate}\ \ D_L.
    \eea
Remember $D_R+D_L=1$. Generally, for $n$ particles existing at a
common site $x$, one finds from (\ref{4}):
 \bea\label{22}
 {\partial \over \partial
 t}P(\underbrace{x,\cdots,x}_n)&=&
 D_RP(x-1,\underbrace{x,\cdots,x}_{n-1})+
 D_LP(\underbrace{x,\cdots,x}_{n-1},x+1)\cr
 &&+\sum_{j=2}^n
 D_RP(x_1=x,\cdots,x_{j-1}=x,x_j=x-1,x_{j+1}=x\cdots,x_n=x)\cr
 &&+\sum_{j=2}^nD_LP(x_1=x,\cdots,x_{j-2}=x,x_{j-1}=x+1,x_j=x,\cdots,x_n=x)\cr
  && -n P(\underbrace{x,\cdots,x}_n)\cr&=&
  D_RP(x-1,\underbrace{x,\cdots,x}_{n-1})+
 D_LP(\underbrace{x,\cdots,x}_{n-1},x+1)-
 P(\underbrace{x,\cdots,x}_n),
  \eea
which is obviously the evolution equation of reactions (\ref{7}).
\section{Totally asymmetric diffusion with $D_L=0$}
In $D_L=0$, the boundary condition (\ref{3}) and (\ref{6}) becomes
 \bea\label{23}
 P(x,x-1)&=&P(x-1,x-1), \cr
 P(x,x-1)&=&P(x,x),
 \eea
respectively, which describe the interactions
  \be\label{24}
  mn \rightarrow m-l,n+l \
\ \ \ {\rm with
 \ rate}\ \ 1
  \ee
and
  \be\label{25}
  mn \rightarrow m-1,n+1 \
\ \ \ {\rm with
 \ rate}\ \ 1,
  \ee
respectively. But as was first noted in \cite{20} for ASEP, the
linear combination of these two boundary conditions may result a
new integrable model. So we consider
  \be\label{26}
  P(x,x-1)=\l P(x,x)+\m P(x-1,x-1),
  \ee
as the boundary condition for the following master equation
 \be\label{27}
 {\partial \over \partial
 t}P(x_1,\cdots,x_N;t)=\sum_{j=1}^N\left[
 P(x_1,\cdots,x_{j-1},x_j-1,x_{j+1}\cdots,x_N;t)-
 P(x_1,\cdots,x_N;t)\right].
  \ee
First we note that
   \be\label{28}
  {\partial \o {\partial t}}\s_{x_2}\s_{x_1\leq x_2}P(x_1,x_2;t)=
  ( \l+\m-1)\s_xP(x,x;t),
  \ee
from which the conservation of number of particles results
  \be\label{29}
  \l+\m=1.
  \ee
To obtain the resulting reactions, besides the diffusion
$10\rightarrow 01$ with rate 1, we first consider ${\dot P(x,x)}$.
Using (\ref{26}) and (\ref{27}), it is found
 \be\label{30}
 {\dot P}(x,x)=P(x-1,x)+\m P(x-1,x-1)
 -(2-\l )P(x,x),
 \ee
which represents the reaction $20\rightarrow 02$, with rate $\m
=1-\l$, as the source and sink of this state. To find the
reactions in general case, we first prove a lemma.\\
{\bf Lemma:} Equation (\ref{26}) implies, for arbitrary $n$, the
following
  \be\label{31}
  P(\underbrace{x,\cdots,x}_{n},x-1)=r_{n+1} P(\underbrace{x-1,\cdots,x-1}_{n+1})
  +(1-r_{n+1})P(\underbrace{x,\cdots,x}_{n+1}),
  \ee
with $r_n$ defined through (\ref{10}).\\
{\bf Proof:} We proceed by induction. For $n=1$, (\ref{31})
reduces to (\ref{26}). Assuming (\ref{31}) is correct for $n-1$,
then using (\ref{26}), we have
 \bea\label{32}
 P(\underbrace{x,\cdots,x}_{n},x-1)&=&\l
 P(\underbrace{x,\cdots,x}_{n+1})
  +\m P(\underbrace{x,\cdots,x}_{n-1},x-1,x-1)\cr
 &=&\l P(\underbrace{x,\cdots,x}_{n+1})+\m [
  r_{n} P(\underbrace{x-1,\cdots,x-1}_{n+1})
  +(1-r_{n})P(\underbrace{x,\cdots,x}_{n},x-1)],\cr &&
  \eea
so
  \be\label{33}
  P(\underbrace{x,\cdots,x}_{n},x-1)=r_{n+1} P(\underbrace{x-1,\cdots,x-1}_{n+1})
  +s_{n+1}P(\underbrace{x,\cdots,x}_{n+1}),
  \ee
where
  \be\label{34}
  r_{n+1}={\m r_n \o {1-\m (1-r_n)}} \ \ , \ \
  s_{n+1}={\l  \o {1-\m (1-r_n)}}.
  \ee
It is seen that $r_{n+1}+s_{n+1}=1$. The value of $r_n$ can be
found by solving the first equation of (\ref{34}), which can be
written as
  \be\label{35}
  r^{-1}_{n+1}=1+{\l\o\m}r^{-1}_n.
  \ee
Using $r_2^{-1}={1\o\m}=1+{\l\o\m}$, (\ref{35}) leads to
eq.(\ref{10}) for $r_n$s. This proves the lemma.$\blacksquare$

We now consider the evolution of $P(\underbrace{x,\cdots,x}_{n})$.
Using (\ref{27}) and (\r{31}), we find
 \bea\label{36}
 {\partial \over \partial
 t}P(\underbrace{x,\cdots,x}_n)&=&
 \sum_{j=1}^n
 P(x_1=x,\cdots,x_{j-1}=x,x_j=x-1,x_{j+1}=x\cdots,x_n=x)
 -n P(\underbrace{x,\cdots,x}_n)\cr
 &=& \sum_{j=1}^n
 [ r_jP(x_1=x-1,\cdots,x_j=x-1,x_{j+1}=x\cdots,x_n=x)
 +(1-r_j) P(\underbrace{x,\cdots,x}_n)]\cr &&
 -n P(\underbrace{x,\cdots,x}_n)\cr
  &=& \sum_{j=1}^n
  r_jP(x_1=x-1,\cdots,x_j=x-1,x_{j+1}=x\cdots,x_n=x)
  -(\s_{j=1}^nr_j) P(\underbrace{x,\cdots,x}_n).\cr &&
  \eea
This equation shows that a collection of $j$ particles hops from
site $x-1$ to site $x$ with rate $r_j$, and this proves that the
induced reactions of the boundary condition (\r{8}) are those
indicated in (\r{9}).

At $\l =1$ ($\m =0$), eq.(\r{10}) results $r_l=\delta_{l1}$. So
the reactions are
  \be\label{37}
  mn \rightarrow m-1,n+1,  \ \ \ \ {\rm with \ rate} \ 1.
  \ee
This is nothing but eq.(\r{25}). At $\l =0$ ($\m =1$), eq.(\r{10})
gives $r_l=1$ for all $l$s. So the reactions are those shown in
(\r{24}). In fact, at $\l=0$, all the multi-particle hoppings
occur with equal rate 1. By increasing $\l$, the greater number of
simultaneous hoppings happen with lower rates, until at $\l=1$, in
which only the one-particle hopping is allowed.
\section{Annihilation-diffusion processes}
Adding annihilation to the previous reactions, results in the
decreasing of the number of particles with time. Therefore the
parameters $\m',$ $\l'$ and ($\l,$ $\m$) in eqs.(\r{14}), (\r{15})
and (\r{26}) become
  \bea\label{38}
  \m'>1&&  ({\rm for \ boundary \ condition\ (\ref{14})}),\cr
 \l'<1&&  ({\rm for \ boundary \ condition\ (\ref{15})}), \cr
 \m+\l<1&&  ({\rm for \ boundary \ condition\ (\ref{26})}).
  \eea
These are obtained from eqs.(\r{17}), (\r{18}) and (\r{28}),
respectively. Note that the annihilation interactions only appear
in sink terms of the master equation, since if we consider an
initial state with $n$ particles, no annihilation process can lead
to a $n$-particle state at any later time $t$.

To obtain the corresponding interactions of either of boundary
conditions, we begin with (\r{15}) with $\l'<1$. Considering
${\dot P}(x,x)$, one finds
 \be\label{39}
 {\dot P}(x,x)=D_RP(x-1,x)+D_LP(x,x+1)
 -(2-\l')P(x,x),
 \ee
which shows the annihilation rate $1-\l'$ for the state of two
particle at a common site. In general case, one finds, similar to
eq.(\r{22}):
 \be\label{40}
 {\partial \over \partial
 t}P(\underbrace{x,\cdots,x}_n)=
  D_RP(x-1,\underbrace{x,\cdots,x}_{n-1})+
 D_LP(\underbrace{x,\cdots,x}_{n-1},x+1)-[n-(n-1)\l']
 P(\underbrace{x,\cdots,x}_n),
  \ee
which is the evolution equation of the following reactions:
  \bea\label{41}
  mn &\rightarrow &m-1,n+1 \
\ \ \ {\rm with
 \ rate}\ \ D_R,\cr
 mn &\rightarrow &m+1,n-1 \
\ \ \ {\rm with
 \ rate}\ \ D_L,\cr
n &\rightarrow & \left\
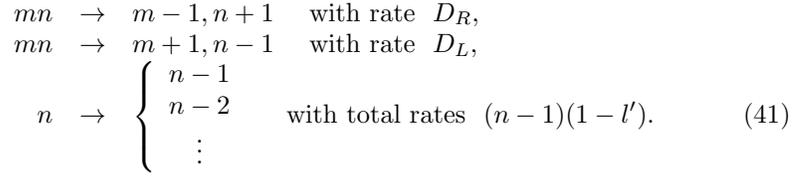
{
\begin{array}{c}
n-1 \\n-2\\ \vdots
\end{array}
\right.
 \ \ \ \ {\rm with \ total
 \ rates}\ \ (n-1)(1-\l').
    \eea
In the case of boundary condition (\r{26}) with $\l+\m<1$, it is
not easy to obtain a compact form for interaction rates in general
case. For example in two- and three-particle sectors, considering
${\dot P}(x,x)$ and ${\dot P}(x,x,x)$, one finds
 \bea\label{42}
  mn &\rightarrow &m-1,n+1 \
\ \ \ {\rm with
 \ rate}\ \ 1,\cr
 mn &\rightarrow &m-2,n+2 \
\ \ \ {\rm with
 \ rate}\ \ \m,\cr
  mn &\rightarrow &m-3,n+3 \
\ \ \ {\rm with
 \ rate}\ \ {\m^2\o{1-\l\m}},\cr
  2 &\rightarrow &1 \
\ \ \ \ \ \ \ \ \ \ \ \ \ \ \ \ \   {\rm with
 \ rate}\ \ \a=1-(\l+\m),\cr
3 &\rightarrow & \left\
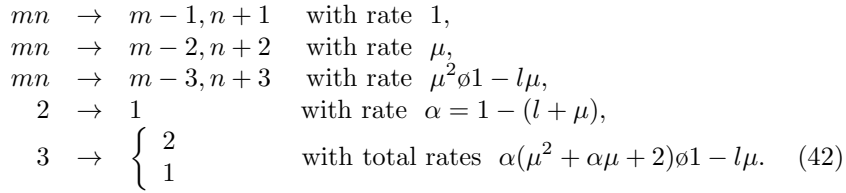
{
\begin{array}{c}
2 \\1
\end{array}
\right.
 \ \ \ \ \ \ \ \ \ \ \ \ {\rm with \ total
 \ rates}\ \ {\a(\m^2+\a\m+2)\o{1-\l\m}}.
    \eea
In this case, we have a two-parameter family of interactions.

When we consider the boundary condition (\r{14}) with $\m'>1$,
again we can not find the compact relations and it is better to
restrict ourselves to the subset of totally asymmetric reactions
with $D_L=0$. This means that we take
 \be\label{43}
 P(x,x-1)=\m P(x-1,x-1) \ \ \ (\m <1)
 \ee
as the boundary condition ( $\m=1/\m'$ ). Considering the general
case $P(x,\cdots,x)$, results
 \bea\label{44}
 {\partial \over \partial
 t}P(\underbrace{x,\cdots,x}_n)&=&
 \sum_{j=1}^n
 P(x_1=x,\cdots,x_{j-1}=x,x_j=x-1,x_{j+1}=x\cdots,x_n=x)
 -n P(\underbrace{x,\cdots,x}_n)\cr
  &=& \sum_{j=1}^n
  \m^{j-1}P(x_1=x-1,\cdots,x_j=x-1,x_{j+1}=x\cdots,x_n=x)
  -n P(\underbrace{x,\cdots,x}_n).\cr &&
  \eea
This equation shows that the sources are the simultaneous hoppings
of $j$ particles ($j=1,\cdots,n$) from the common site $x-1$ to
$x$, with rates $\m^{j-1}$. So the sinks are also these hoppings
and the remaining rate of the sink terms in eq.(\r{44}), i.e.
$n-\s_{j=0}^{n-1}\m^j$, is due to the annihilations of these $n$
particles at site $x$. So, it is proved that the boundary
condition (\r{43}) induces the reactions (\r{12}).
\section{Bethe ansatz solution}
Now we try to solve the resulting evolution equations, in all the
discussed cases, by the following Bethe ansatz
  \be\label{45}
  P({\mathbf x};t)=e^{-E_Nt}\psi({\mathbf x}),
  \ee
where ${\mathbf x}=(x_1,\cdots,x_N)$, and
  \be\label{46}
  \psi({\mathbf x})=\s_\sigma A_\sigma e^{i\sigma ({\mathbf p}).{\mathbf
  x}}.
  \ee
The summation runs over the elements of permutation group of $N$
object. Inserting (\r{46}) in master equation (\r{4}), results
  \be\label{47}
  E_N=\s_{k=1}^N\left( 1-D_Re^{-ip_k}-D_Le^{ip_k}\right).
  \ee
To determine $A_\sigma$, we must insert the expression (\r{46}) in
the boundary conditions. For type 2 model, for instance, we must
use eq.(\r{6}), which results
 \be\label{48}
  D_R\psi(\cdots,x_i=x,x_{i+1}=x-1,\cdots)+
  D_L\psi(\cdots,x_i=x+1,x_{i+1}=x,\cdots)=
  \psi(\cdots,x_i=x,x_{i+1}=x,\cdots),
  \ee
and using (\r{46}), gives
  \be\label{49}
  \left[ D_Re^{-i\sigma (p_{k+1})}+D_Le^{i\sigma (p_{k})}-1
  \right] A_\sigma +
  \left[ D_Re^{-i\sigma (p_{k})}+D_Le^{i\sigma (p_{k+1})}-1
  \right] A_{\sigma\sigma_k}=0.
  \ee
$\sigma_k$ is an element of permutation group which only
interchanges $p_k$ and $p_{k+1}$,
  \be\label{50}
  \sigma_k:(p_1,\cdots,p_k,p_{k+1},\cdots,p_N)\rightarrow
  (p_1,\cdots,p_{k+1},p_{k},\cdots,p_N).
  \ee
Eq.(\r{49}) gives $A_{\sigma\sigma_k}$ in terms of $A_\sigma$ as
following
  \be\label{51}
  A_{\sigma\sigma_k}=S^{(2)}( \sigma (p_{k}),
  \sigma (p_{k+1}))A_\sigma
  \ee
where
 \be\label{52}
  S^{(2)}(z_1,z_2)=-{{D_Rz_2+D_Lz_1^{-1}-1}\o{D_Rz_1+D_Lz_2^{-1}-1}},
  \ee
where $z_k=e^{-ip_k}$. Eq.(\r{51}) allows one to compute all
$A_\sigma$s in terms of $A_1$, which is set to unity.

The same procedure can be applied to other boundary conditions.
For example for type 3 model, boundary condition (\r{8}), one
finds
  \be\label{53}
  S^{(3)}(z_1,z_2)=-{{\l+\m(z_1+z_2)-z_2}\o{\l+\m(z_1+z_2)-z_1}},
  \ee
and for boundary condition (\r{15}), with $\l'<1$,
 \be\label{54}
  S(z_1,z_2)=-{{D_Rz_2+D_Lz_1^{-1}-\l'}\o{D_Rz_1+D_Lz_2^{-1}-\l'}}.
  \ee
These solutions can be used, in principle, to calculate the
conditional probabilities $P(x_1,\cdots,x_N;t|y_1,\cdots,y_N;0).$
This is the probability of finding the particles at time $t$ at
sites $x_1,\cdots,x_N$, if at $t=0$, they were at sites
$y_1,\cdots,y_N$, respectively. But unfortunately, the standard
method, used for example in \cite{19,20,21,22,23}, can not be used
here. This is because the initial condition
  \be\label{55}
  P({\mathbf x};0|{\mathbf y};0)=\delta_{{\mathbf x},{\mathbf y}},
  \ee
is satisfied by standard expression
  \be\label{56}
  P({\mathbf x};t|{\mathbf y};0)=\int {d^Np\o
  (2\pi)^N}e^{-E_Nt}e^{-i{\mathbf p}.{\mathbf y}}\psi({\mathbf
  x}),
  \ee
only when $y_i<y_{i+1}$ and $x_i<x_{i+1}$. This condition only
satisfied by exclusion processes and for processes without
exclusion, we must look for other methods.
\section{$N$-particle conditional probabilities}
In some special cases, it is possible to calculate the conditional
probabilities in terms of a specific determinant. This was first
proved in \cite{19}, and then used  for other cases in
\cite{20,21}. Now we want to check that this method does work
here, and as a specific example, we consider the type 4 model with
the boundary condition (\r{11}).

We set the following ansatz for $N$-particle conditional
probabilities
  \be\label{57}
  P({\mathbf x};t|{\mathbf y};0)=e^{-Nt}\det[G ({\mathbf x};t|{\mathbf
  y};0)],
  \ee
where $G$ is a $N\times N$ matrix with elements
  \be\label{58}
  G_{ij}({\mathbf x};t|{\mathbf y};0) = g_{i-j}(x_i -y_j;t).
  \ee
Inserting (\r{57}) in eqs.(\r{27}) and (\r{11}), results
  \bea\label{59}
  {\partial\o\partial t}G_i(x;t)&=&G_i(x-1;t),\cr G_{i-1}(x;t)
  &=& \m G_{i-1}(x-1;t) +\b G_i(x-1;t),
  \eea
where $G_i$ denotes the $i$-th row of matrix $G$, and $\b$ is an
arbitrary constant. In terms of functions $g_p(n;t)$, (\r{59})
becomes
  \bea\label{60}
  {\partial\o\partial t}g_p(n;t)&=&g_p(n-1;t),\cr
  g_p(n;t)&=&\m g_p(n-1;t)+\b g_{p+1}(n-1;t).
  \eea
Introducing the $z$-transform
  \be\label{61}
  {\tilde g}_p(z,t)=\s_{n=-\infty}^\infty z^ng_p(n;t),
  \ee
we find
  \bea\label{62}
  {\partial\o\partial t}{\tilde g}_p(z,t)=
  z{\tilde g}_p(z,t)&\Rightarrow &{\tilde g}_p(z,t)
  =e^{zt} {\tilde g}_p(z,0),\cr {\tilde g}_{p+1}(z,t)
  &=&{1-\m z\o \b z}{\tilde g}_p(z,t).
  \eea
The second equation yields
  \be\label{63}
  {\tilde g}_{p}(z,0)=\left( {1-\m z\o \b z}\right)^p{\tilde g}_0(z,0).
  \ee
Using $P(x;0|y;0)=g_0(x-y;0)=\delta_{x,y}$, one finds ${\tilde
g}_0(z,0)=1$. Finally
  \be\label{64}
  {\tilde g}_{p}(z,t)=e^{zt}\left( {1-\m z\o z}\right)^p,
  \ee
in which we choose $\b=1$. The functions $g_p(n;t)$ can be
obtained by expanding the generating functions ${\tilde
g}_{p}(z,t)$. For $p\geq 0$, the expansion yields
  \be\label{65}
  g_{p}(n;t)=\s_{m=0}^p
  \left( \begin{array}{c} p\\ m\\ \end{array} \right)
   (-\m)^m {t^{n+p-m}\o (n+p-m)!},
   \ee
and for negative $p$, it yields
   \be\label{66}
  g_{-|p|}(n;t)=\s_{m=0}^{n-|p|}
  \left( \begin{array}{c} |p|+m-1\\ m\\ \end{array} \right)
   \m^m {t^{n-|p|-m}\o (n-|p|-m)!}.
   \ee
We have thus obtained the explicit relation for conditional
probabilities.

It can be checked that the resulting function satisfies the
desired initial condition. At $t=0$, where $x_i=y_i$, we have
$g_{-|p|}(0;t)=0$ and $g_0(n;0)=\delta_{n,0}$, which result
$P({\mathbf x};0|{\mathbf x};0)=1$. Also it may be interesting to
obtain the rate of decay of a delta function distribution. Suppose
at $t=0$, there are $N$ particles at the same site $y$. We want to
obtain the probability of finding all the particles at their
initial positions at later time $t$, i.e. $P({\mathbf
y};t|{\mathbf y};0)$, where ${\mathbf y}=(y,\cdots,y)$. Using
$g_{-|p|}(0;t)=0$ and $g_0(0;t)=1$, we obtain
  \bea\label{67}
  P({\mathbf y};t|{\mathbf y};0)&=&e^{-Nt}\det[G ({\mathbf y};t|{\mathbf
  y};0)]\cr &=&e^{-Nt},
  \eea
which is independent of $\m$! It is an interesting result. At
$\m=1$, where there is no annihilation, the rate of simultaneous
hoppings of particles are all 1, irrespective of the number of
particles. By decreasing $\m$, the rates of simultaneous hoppings
decrease with increasing the number of particles (see
eq.(\r{12})). At the same time, the rate of annihilation increases
in such a way that the total decaying rate remains constant.

The above determinant method can be applied to any model which its
boundary condition equation contains only two terms. Otherwise it
does not lead to the consistent relations for matrix elements
$g_p(n;t)$'s. So for $D_R=0$ {\it or} $D_L=0$ cases of boundary
conditions (\ref{3}) and (\ref{6}), and $\lambda =0$ {\it or} $\mu
=0$ cases of boundary condition (\ref{8}), this method leads to
explicit expressions for $N$-particle conditional probabilities.
\section{Conclusion}
In the previous sections, we use all the allowed generalization of
boundary condition in the asymmetric reaction-diffusion processes
without exclusion, to obtain several new integrable models for
one-species cases. The further natural generalization , which has
been discussed in many papers for the ASEP case, is the
multi-species extensions of these models. In multi-species
studies, one considers a $p$-species system with particles
$A_1,\cdots,A_p$. The basic objects are the probabilities
$P_{\a_1\cdots\a_N}(x_1,\cdots,x_N;t)$ for finding at time $t$ the
particles of type $\a_1$ at $x_1$, particle of type $\a_2$ at
$x_2$, etc. The master equation, instead of (\r{4}), is \cite{23}
 \bea\label{68}
 {\partial \over \partial
 t}P_{\a_1\cdots\a_N}(x_1,\cdots,x_j,\cdots,x_N;t)&=&\sum_{j=1}^N
 D_RP_{\a_1\cdots\a_N}(x_1,\cdots,x_{j-1},x_j-1,x_{j+1}\cdots,x_N;t)\cr
 &&+\sum_{j=1}^ND_LP_{\a_1\cdots\a_N}(x_1,\cdots,x_{j-1},x_j+1,x_{j+1}\cdots,x_N;t)\cr
  && -NP_{\a_1\cdots\a_N}(x_1,\cdots,x_j,\cdots,x_N;t).
  \eea
Now if we want to use this equation for reactions without
exclusion, the problem will arise when some of the particles are
in the same position. Consider, for example, the two-species case
$P_{\a_1\a_2}(x,x)$. It is seen that eq.(\r{68}) has the term
$P_{\a_1\a_2}(x-1,x)$ as the source term of the desired state, but
does not contain $P_{\a_2\a_1}(x-1,x)$, which is as important as
the first term. In fact, this is the source of many difficulties
arise in multi-species extension of reaction-diffusion processes
which have not exclusion. So it seems that the integrable models
discussed in the previous sections are {\it all} that one can
obtain in this context.

{\bf Acknowledgement:}  I would like to thank the research council
of
 the University of Tehran for partial financial support.
%\newpage


\begin{thebibliography}{9}
\bibitem{1} V. Privman (editor), {\it Nonequilibrium Statistical
Mechanics in One Dimension} (Cambridge University Press,
Cambridge, 1977).
\bibitem{2} H. Spohn: {\it Large Scale Dynamics of Interacting
Particles} (Springer, 1991).
\bibitem{3} S. Katz, J. L. Lebowitz, \& H. Spohn; J. Stat. Phys. {\bf 34} (1984) 497.
\bibitem{4} L. Garrido, J. Lebowitz, C. Maes, \& H. Spohn; Phys. Rev.
{\bf A42} (1990) 1954.
\bibitem{5} L. P. Kadanoff \& J. Swift; Phys. Rev. {\bf 165} (1968) 310.
\bibitem{6} T. M. Ligget: {\it Interacting Particle Systems } (Springer-Verlag, New York,, 1985).
\bibitem{7} B. Derrida; Phys. Rep. {\bf 301} (1998) 65.
\bibitem {8} C. T. MacDonald, J. H. Gibbs, \& A. C. Pipkin; Biopolymers {\bf 6} (1968) 1.
\bibitem {9} G. M. Sch\"{u}tz; Int. J. Mod. Phys. {\bf B11}
(1997) 197.
\bibitem {10} K. Nagel; Phys. Rev. {\bf E53} (1996) 4655.
\bibitem {11} J. Krug \& H. Spohn: {\it Solids far from equilibrium }, ed. C. Godreche ( Cambridge University
Press, Cambridge, 1991).
\bibitem {12} T. Halpin-Healey \& Y. -C. Zhang; Phys. Rep. {\bf
254} (1995) 215.
\bibitem {13} J. M. Burgers: {\it The nonlinear diffusion equation }
( Reidel, Boston, 1974).
\bibitem {14} B. Derrida, S. A. Janowsky, J. L. Lebowitz, \& E. R. Speer; Europhys. Lett. {\bf 22} (1993) 651.
\bibitem {15} P. A. Ferrari \& L. R. G. Fontes; Probability Theory Related Fields {\bf 99} (1994) 305.
\bibitem{116} R. Bundschuh; Phys. Rev. {\bf E65} (2002) 031911.
\bibitem{117} R. Lipowsky, S. Klumpp, \& T. M. Nieuwenhuizen; Phys. Rev. Lett. {\bf 87} (2001) 108101.
\bibitem {16} D. ben-Avraham \& S. Havlin: {\it Diffusion and
reactions in fractal and disordered systems} ( Cambridge
University Press, Camdridge, 2000).
\bibitem {17} G. M. Sch\"{u}tz, in: {\it Phase transitions and critical
phenomena }(C. Domb and J. Lebowitz Eds.), Vol. 19, Academic Press
(London, 2000).
\bibitem {18} M. Henkel, in: {\it Classical and quantum nonlinear integrable systems: theory and
application}, edited by A. Kundu, IOP Press (Bristol, 2003) pp.
256-287 (cond-mat/0303512).
\bibitem {19} G. M. Sch\"{u}tz; J. Stat. Phys. {\bf 88} (1997) 427.
\bibitem {20} M. Alimohammadi, V. Karimipour, \& M. Khorrami; Phys. Rev. {\bf E57} (1998) 6370.
\bibitem {21} M. Alimohammadi, V. Karimipour, \& M. Khorrami; J. Stat. Phys. {\bf 97} (1999) 373.
\bibitem {22} M. Alimohammadi; Eur. Phys. J. {\bf B42} (2004) 415.
\bibitem{23} M. Alimohammadi \& Y. Naimi; J. Math. Phys. {\bf 46}
(2005) 053306.
\bibitem{24} N. M. Bogoliubov \& T. Nassar; Phys. Lett. {\bf A234}
(1997) 345.
\bibitem{25} T. Sasamoto \& M. Wadati; J. Phys. {\bf A31} (1998) 6057.
\bibitem{26} T. Sasamoto \& M. Wadati; Phys. Rev. {\bf E58} (1998) 4181.
\end{thebibliography}
\end{document}